\documentstyle[prl,aps,twocolumn,epsf]{revtex}
\begin{document}


\title{Induced Anticlinic Ordering and Nanophase Segregation of Bow-Shaped
Molecules in a Smectic Solvent}
\author{Prabal K. Maiti, Yves Lansac, Matthew A. Glaser and Noel A. Clark}
\address{Department of Physics and \\
Ferroelectric Liquid Crystal Material Research Center,\\
University of Colorado, Boulder, CO 80309}
\address{
\centering{
\medskip\em
{}~\\
\begin{minipage}{14cm}
Recent experiments indicate that doping low concentrations of 
bent-core molecules into calamitic smectic solvents can induce anticlinic 
and biaxial smectic phases. We have carried out Monte Carlo (MC) simulations of
mixtures of rodlike molecules (hard spherocylinders with length/breadth ratio
$L_{\rm rod}/D = 5$)  and bow- or banana-shaped molecules (hard spherocylinder dimers 
with length/breadth ratio $L_{\rm ban}/D = 5$ or $2.5$ and opening angle $\psi$) 
to probe the molecular-scale organization and phase behavior of rod/banana
mixtures. We find that a low concentration ($3\%$) of $L_{\rm ban}/D = 5$ dimers
induces anticlinic (SmC$_A$) ordering in an untilted smectic (SmA) phase for 
$100^\circ \le \psi < 150^\circ$. 
For smaller $\psi$, half of each bow-shaped molecule is nanophase segregated 
between smectic layers, and the smectic layers are untilted. For $L_{\rm ban}/D = 2.5$,
no tilted phases are induced. However, with decreasing $\psi$ we observe a sharp 
transition from {\sl intralamellar} nanophase segregation (bow-shaped molecules
segregated within smectic layers) to {\sl interlamellar} nanophase segregation
(bow-shaped molecules concentrated between smectic layers) near $\psi = 130^\circ$.
These results demonstrate that purely entropic effects can lead to surprisingly 
complex behavior in rod/banana mixtures.
{}~\\
\end{minipage}
}}

\maketitle

Liquid crystals (LCs) are exotic solvents that impose 
orientational and/or positional ordering on solutes. 
Conversely, LC ordering and properties are highly sensitive to the 
influence of low-symmetry solutes.
For example, 
if a chiral solute is added to a nematic a uniform helical twist of the 
director is induced \cite{degenes}. By properly choosing the 
chiral solute the helical twisting power and hence the helical pitch
can be controlled and this has potential application in
electro-optic displays. If the starting achiral mesophase is SmC, chiral solutes
induce the chiral SmC$^\ast$ phase which is ferroelectric \cite{stege,stege1}. 
The induced ferroelectric SmC$^\ast$ has variety of technological 
applications \cite{sven}.
Addition of organic solvent in SmA and SmC phase can increase the smectic layer 
spacing in which solvent is intercalated between smectic layers \cite{rieker}.
Also the smectic ordering can significantly enhance polymerization rate in a 
monomer-LC mixtures \cite{Guymon}. 
Results from previous computer simulation studies showed that trans-cis photoisomerization 
of azo-benzene based LC's in smectic solvent leads to intralamellar nanophase
segregation \cite{lansac}. All these phenomena clearly demonstrate that by appropriately
changing the solute  or the liquid crystal solvent, material properties can be manipulated
for specific technological application. 

In recent years a new class of smectic liquid crystal (SmCP phase)
made up of achiral molecules with bent cores has been discovered
\cite{niori,link}. The molecules arrange into 
macroscopically chiral layers 
characterized by polar ordering and molecular tilt about layer normal,
producing spontaneous polarization. 
Recent experimental results indicate that a very low concentration
of bow-shaped molecules in a smectic solvent can give rise to biaxial SmA/tilted
phase \cite{pratibha}. It has also been shown that the polar properties of 
some liquid crystals can be changed from ferroelectric (FE) to antiferroelectric
(AF) by doping with bow-shaped molecules \cite{gorecka}. 
However, a clear understanding of all these 
phenomena at a molecular level, essential for molecular engineering is lacking. 
For example, to select the suitable solutes shape to obtain 
desired macroscopic properties. Computer simulation can help us to develop a molecular
scale interpretation of experimental results and provide theoretical guidance
for further experimental studies. In simulation we can gradually change the molecular shape 
of the solute or solvent and see how these changes affect the large-scale order of 
liquid crystal systems.

To investigate the effect of low concentration of bow-shaped molecules in a 
smectic solvent, we have carried out a series of 
Monte Carlo (MC) simulation in a mixture of rodlike and bow- or banana-shaped molecules 
Some of the questions we want to address
are: what new LC phase can be induced by a low concentration of bow-shaped 
molecules in a calamitic LC solvent; how are the bow-shaped molecules are 
distributed in a LC solvent? How is the structure of the solvent is modified?
Under what conditions does nanophase segregation occur. Our model simulations 
give valuable insight into some of these questions. To a fair approximation 
the banana molecules with small bent angle can be thought of as behaving like
platelike molecules. In that respect our simulation can shed light on the 
phase behavior of such mixture.

\begin{figure}[h]
\epsfxsize=1.7 in
\centerline{\epsffile{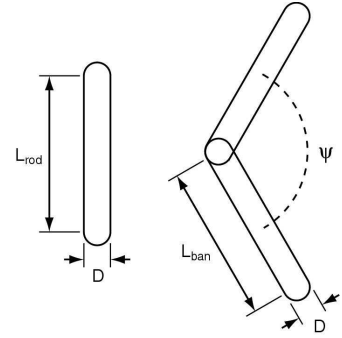}}
\caption{\protect{A schematic representation of model system.  }}
\label{modelfig}
\end{figure}

The smectic solvent is made up of rodlike molecules represented by 
hard spherocylinders of length/breadth ratio $L_{\rm rod}/D$. The bow-shaped 
molecules are modeled as spherocylinder dimers of length/breadth ratio
$L_{\rm ban}/D$ with opening angle $\psi$ (figure \ref{modelfig}). The
model of bow-shaped molecules as spherocylinder dimers has been 
investigated through computer simulation and gives rise to rich phase 
behavior \cite{camp1,lansac1} and captures some of the essential 
physics of such liquid crystalline system. The idea behind using such 
hard core model is that liquid crystal phase behavior is largely 
entropy driven and determined by the hard core repulsion between the liquid 
crystal mesogens. Hard spherocylinder provides a simple model both in terms 
of computational ease and theoretical approach.

We have performed MC simulations of rod/banana mixtures under
$NPT$ (constant particle number, constant pressure,
and constant temperature) ensemble with periodic boundary conditions.
The simulated systems contained
$N = 400$ molecules in a orthorhombic box. The number of rodlike molecules 
was $N_{\rm rod} = 388$ and the number of
banana molecules was $N_{\rm ban} = 12$. The concentration of banana molecules,
defined as $c_{\rm ban} = N_{\rm ban}/N$, is $3\%$. 
Initially, the system was prepared in a SmA state at high pressure.
Each trial MC move consisted of a random displacement and reorientation of a 
randomly selected rod or banana molecule, with the move accepted according to 
a Metropolis criterion \cite{allen}. 
Simulations were carried out for two sets of length to breadth ratios for rods and bananas. 
In the first case, $L_{\rm rod}/D = 5$ and $L_{\rm ban}/D = 5$ 
(the overall banana length is twice the rod length), while
in the second case, $L_{\rm rod}/D = 5$ and $L_{\rm ban}/D = 2.5$ 
(the overall banana length is equal to the rod length). 
For convenience we introduce reduced units, in which the reduced pressure $P\ast$ is
defined as $P^\ast = \beta P v_0$ and the reduced density as $\rho^\ast = \rho v_0$, 
where $v_0$ is the volume of a hard spherocylinder of length $L$ and breadth $D$.
For all simulations, the pressure was kept fixed at $P^\ast = 11$. This
pressure corresponds to the middle of the SmA phase for $L/D = 5$ spherocylinders
\cite{macgroth}.

\begin{figure}[h]
\epsfxsize=2.5 in
\centerline{\epsffile{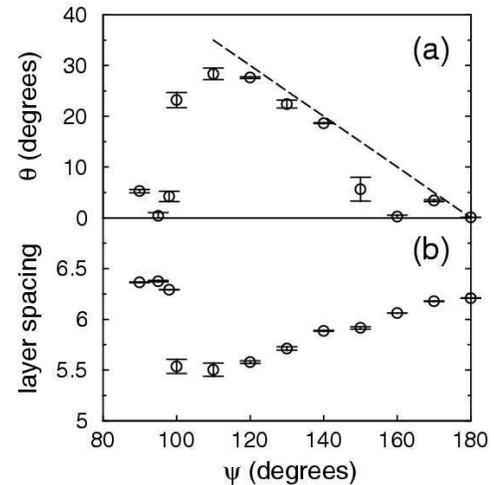}}
\caption{\protect{Induced tilt angle ($\theta$)(a) and Layer 
spacing (b) as a function of opening angle $\psi$ for the case when $L_{\rm rod}/D = 5$ 
and $L_{\rm ban}/D = 5$.}}
\label{tiltangle}
\end{figure}

\begin{figure}[h]
\epsfxsize=2.5 in
\centerline{\epsffile{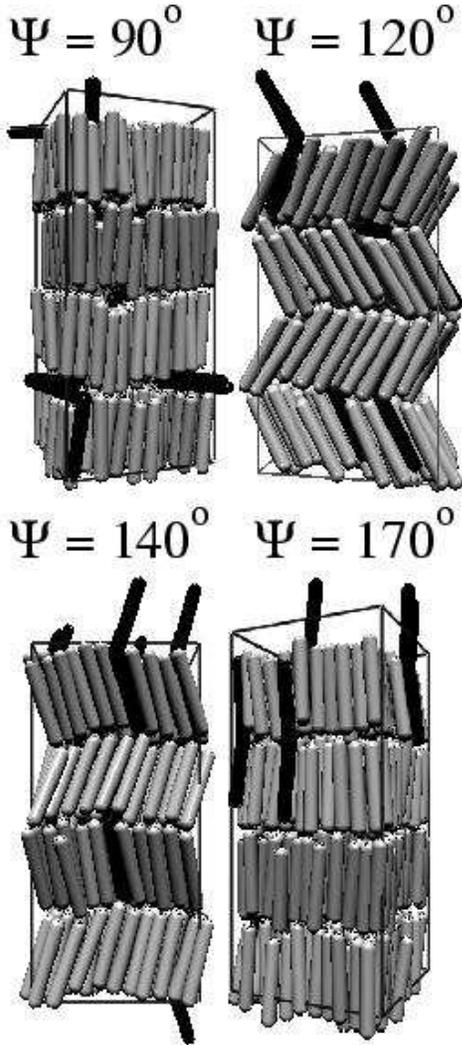}}
\caption{\protect{Equilibrated snapshot of the system for $\psi = 90^\circ, 120^\circ,
140^\circ$ and $170^\circ$ respectively. The 
banana molecules are in black and rods are in gray color. The parameters are same as
in figure \ref{tiltangle}. }}
\label{lsnapshot}
\end{figure}

Starting with an untilted SmA phase, we find that low concentration of 
$L_{\rm ban}/D = 5$ banana molecules induces anticlinic (SmC$_A$) ordering 
in the smectic host for
sufficiently large opening angle $\psi$. This behavior is evident in
Figure~\ref{tiltangle}(a), where we plot the anticlinic tilt angle $\theta$ 
with respect to the layer normal as a function of opening angle $\psi$ 
\cite{anticlinic_tilt}. 
For $100^\circ \le \psi < 150^\circ$, the molecules within each layer exhibit
a uniform tilt, and the system as a whole develops anticlinic ordering.
Over this range of opening angles, the anticlinic tilt angle is approximately
related to the opening angle by $\theta = (\pi - \psi) / 2$ 
(dashed line in Figure~\ref{tiltangle}(a).
For $\psi < 100^\circ$, half of each banana molecule is nanophase segregated
between smectic layers, and the layers are untilted. For 
$\psi > 150^\circ$, the molecules within each layer are uniformly
tilted, but there is no overall anticlinic ordering on average. Thus, we
observe a transition from SmA to SmC$_A$ and back to SmA with decreasing
opening angle. This behavior is apparent in instantaneous configurations 
from the simulations \ref{lsnapshot}. 
In figure \ref{tiltangle}(b) we have plotted the smectic layer spacing vs. opening 
angle $\psi$. The layer spacing decreases with decreasing opening angle, indicative of
increasing molecular tilt, but increases abruptly for $\psi > 100^\circ$. Note that
the layer spacing at $\psi = 90^\circ$ is larger than that at $\psi = 180^\circ$,
owing to the nanophase segregation of half of each banana molecule between smectic
layers.

\begin{figure}[h]
\epsfxsize=2.5 in
\centerline{\epsffile{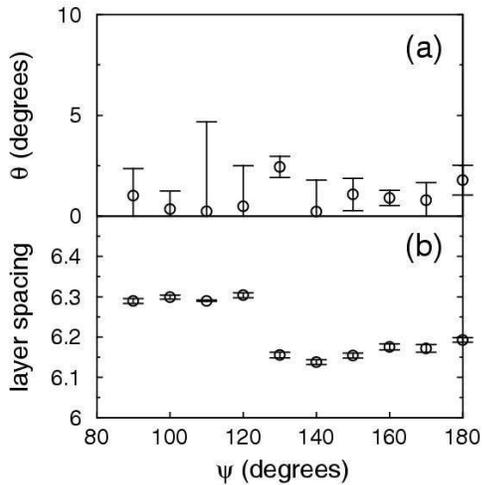}}
\caption{\protect{Tilt angle (a) and smectic layer spacing (b) as a 
function of the opening angle $\psi$ for the case when $L_{\rm rod}/D = 5$
and $L_{\rm ban}/D = 2.5$.}}
\label{spacingl2.5}
\end{figure}

\begin{figure}[h]
\epsfxsize=2.5 in
\centerline{\epsffile{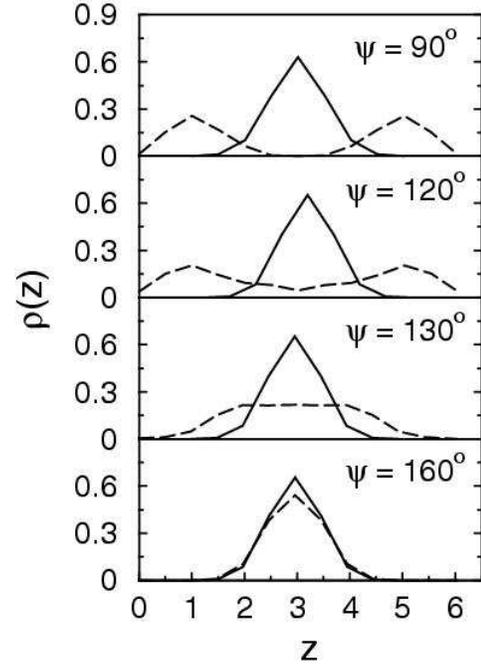}}
\caption{\protect{Density profile for different values of $\psi$ for the case
when $L_{\rm rod}/D = 5$ and $L_{\rm ban}/D = 2.5$. Solid line is for
rods (solvent) and dashed line for bananas. The banana density 
has been scaled by $30$ for clarity.}}
\label{densprof}
\end{figure}

\begin{figure}[h]
\epsfxsize=2.5 in
\centerline{\epsffile{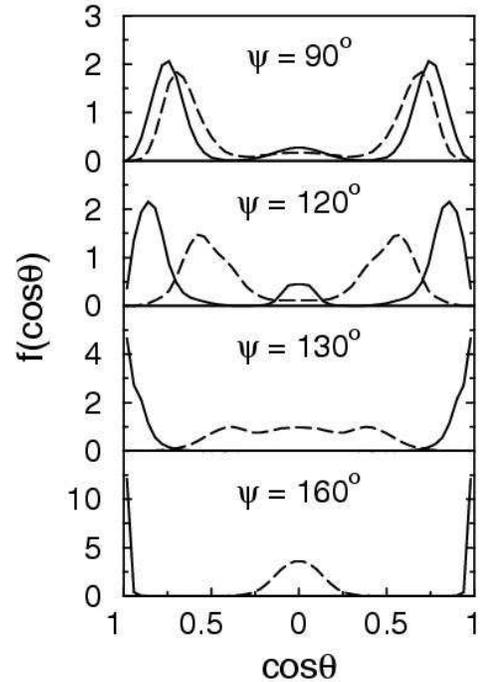}}
\caption{\protect{Orientational distribution of the end-to-end (solid line) and polar 
vector (dashed line) of the banana molecules for different $\psi$.}}
\label{orient_ban}
\end{figure}

\begin{figure}[h]
\epsfxsize=2.5 in
\centerline{\epsffile{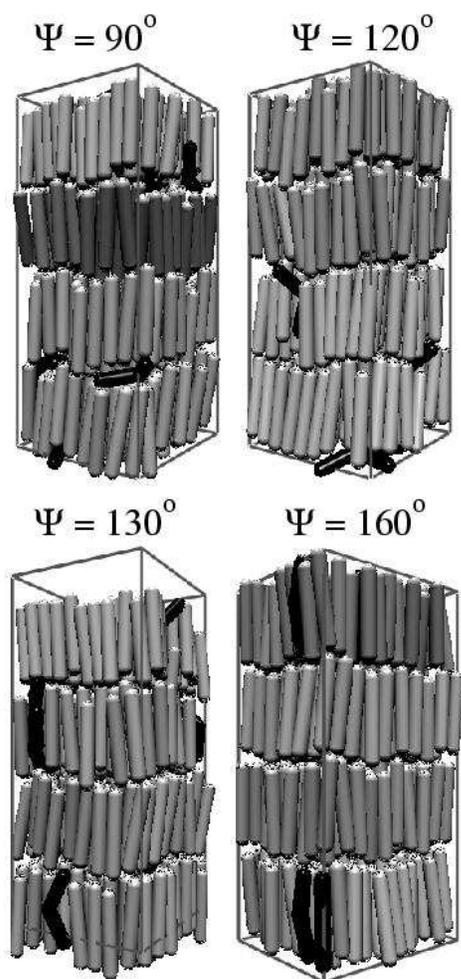}}
\caption{\protect{Equilibrated snapshot of the system for $\psi = 90^\circ$, $120^\circ$,
$130^\circ$ and $160^\circ$ respectively. The 
banana molecules are in black and rods are in gray color.}}
\label{ssnapshot}
\end{figure}

We have also performed simulations of a $3\%$ concentration of $L_{\rm ban}/D = 2.5$ 
banana molecules in an $L_{\rm rod}/D = 5$ spherocylinder host.
In this case no tilted phases are observed (Figure \ref{spacingl2.5}(a)), and 
the system remains in the SmA phase for all $\psi$. 
However, for $\psi \le 130^\circ$ banana density profile (Figure~\ref{densprof}) 
exhibits a deep minima near the center of the smectic layer and two peaks on the 
two sides, which indicate that either half of the banana molecules or the entire
molecule are nanophase
segregated between the smectic layers (interlamellar nanophase segregation). 
The end-to-end vector is preferentially
oriented to the layer normal by $\sim 45^\circ$ with the polar vector 
(unit the vector contained in the plane of the molecule and
passing through the apex of the molecule) orienting
themselves parallel to the layer normal (Figure \ref{orient_ban}). 
For larger $\psi$ there is a transition
from interlamellar to intralamellar nanophase segregation, in which banana molecules 
are segregated within the smectic layers as reflected by both the rod and banana density 
profiles peaked at the center of the smectic layer (Figure~\ref{densprof}.
In this case the end-to-end vector is almost parallel to the layer normal with polar 
vector being in the plane perpendicular to the layer normal (Figure \ref{orient_ban}).
This transition from intralamellar to interlamellar is clearly seen from 
the equilibrated snapshots of the systems shown in figure \ref{ssnapshot}.
Such transition with decreasing opening angle results
in an increase in smectic layer spacing as shown in figure ~\ref{spacingl2.5}(b)).
The same kind of nanophase segregation 
is observed in {\sl trans-cis} photoisomerization of azobenzene in a smectic 
host \cite{lansac}.

Our simulations demonstrate the entropy-driven induction of anticlinic (SmC$_A$)
ordering in a calamitic SmA host by doping with a
low concentration of bow-shaped molecules with overall length twice that of
rodlike solvent molecules. 
The tilted smectic phase may 
be the biaxial smectic phase found in the experiment done by Bangalore group 
in a similar system \cite{pratibha}. 
For smaller $L_{\rm ban}/D$,
we observe an entropy-driven transition from intralamellar to interlamellar 
nanophase segregation of bow-shaped molecules in a mixture of rod and 
banana molecules. In this case, our simulation results demonstrate an
orientational transition as a function of opening angle for banana molecules
in a SmA solvent.
Our model provides a tool which can be useful in 
molecular engineering, for example by selecting suitable shape (in terms of
opening angle and length/breadth ratio) to ensure the desired tilt angle or
layer spacing. Future work will involve a more systematic study 
of the phase behavior and structure of such mixture as a function of
$L_{\rm ban}/D$, $\psi$ and $c_{\rm ban}$, and of the limits of
miscibility of banana molecules in calamitic solvents.

This work was supported by NSF MRSEC Grant DMR 98-09555.

\newpage
\pagebreak

\end{document}